\documentclass[%
 reprint,
preprintnumbers,
 amsmath,amssymb,
 aps, floatfix,]{revtex4-2}

\usepackage{graphicx}
\usepackage{dcolumn}
\usepackage{bm}
\usepackage{xcolor}
\begin{document}

\preprint{APS/123-QED}

\title{Differences in the Resistive and Thermodynamic Properties of the Single Crystalline Chiral Superconductor Candidate SrPtAs}

\author{A. Weiland, F. B. Santos, J. D. Thompson, E. D. Bauer,  S. M. Thomas, and P. F. S. Rosa}
\affiliation{
MPA-Q, Los Alamos National Laboratory, Los Alamos, New Mexico 87545, U.S.A.}
\date{\today}

\begin{abstract}
The locally non-centrosymmetric superconductor SrPtAs is  proposed to host a topological chiral \textit{d}-wave state, but experimental reports have been limited to polycrystalline samples. Here we report the synthesis of single crystalline SrPtAs grown from Pb flux. SrPtAs crystallizes in the hexagonal space group \textit{P}6$_{3}$/\textit{mmc} with lattice parameters \textit{a}~=~4.2445(4)~{\AA} and \textit{c}~=~8.9513(18)~{\AA}. Magnetic susceptibility and electrical resistivity measurements reveal a superconducting transition at T$_c$~$\sim$2.2~K, in agreement with previous reports on polycrystalline samples. Surprisingly, heat capacity data show only a small bulk transition at 0.7 K. We discuss the possible origins of the discrepancy between the various measurements.
 \end{abstract}

\maketitle

SrPtAs, the first hexagonal pnictide-based superconductor \cite{RN80}, is a potential chiral \textit{d}-wave superconductor due to its hexagonal symmetry and quasi-two-dimensional multiband Fermi surface \cite{RN88, RN30}. Chiral superconductors, whose complex superconducting gap function breaks time reversal symmetry (TRS), are of interest due to their non-trivial topology and the potential for quantum computing applications \cite{RN79}. In non $s$-wave pairing, the gap function contains nodes which may be detrimental for superconductivity, but chiral superconducting order parameters may naturally reduce the number of nodes.

In high-symmetry crystal systems, multiple gap functions can be degenerate. For example, in hexagonal systems, $d_{x^2-y^2}$ and $d_{xy}$ superconducting channels are degenerate, and a linear combination of the two results in a chiral $d_{x^2-y^2}$~$\pm$ i$d_{xy}$ state that would maximize the condensation energy of the superconducting state. Superconductors that contain a high-symmetry crystal structure, strong spin-orbit coupling, magnetic interactions, noncentrosymmetry, TRS breaking, and a quasi-two-dimensional Fermi surface are suitable systems to search for chiral superconductivity. Few known materials are considered candidate chiral superconductors. Examples include Sr$_2$RuO$_4$ \cite{SRO}, UPt$_3$ \cite{RN70}, URu$_2$Si$_2$ \cite{RN69}, and potentially SrPtAs \cite{RN87, RN88}.

SrPtAs crystallizes in the hexagonal non-symmorphic space group \textit{P}6$_{3}$/\textit{mmc} (No. 194) \cite{RN9}, wherein Pt and As form a honeycomb lattice. This is in contrast to other pnictide superconductors which form with a square lattice such as LaFePO \cite{RN128}, LaFeAsO \cite{RN130}, LiFeAs \cite{RN148}, NaFeAs \cite{RN16} and ThCr$_2$Si$_2$-type structures like (Ba$_{1-x}$K$_x$)Fe$_2$As$_2$ \cite{RN101}. Although globally centrosymmetric, SrPtAs is locally noncentrosymmetric due to the As-Pt layer breaking inversion symmetry.

Initial measurements, including nuclear magnetic resonance and nuclear quadrupole resonance (NMR/NQR) $T^{-1}_1$ relaxation rates \cite{RN89} argue that SrPtAs is a conventional $s$-wave superconductor because the spin-lattice relaxation rate shows a coherence peak and the Knight shift decreases below T$_c$. Magnetic penetration depth data may be fit to an exponential model and the inferred superfluid density agrees with an isotropic Bardeen–Cooper–Schrieffer (BCS) model \cite{RN93}.

However, muon spin relaxation ($\mu$SR) data showing TRS breaking below the superconducting transition temperature, T$_c$, \cite{RN87} coupled with a recent reevaluation of the NMR data \cite{RN30}, indicate that these data could also correspond to $d$-wave pairing. Although SrPtAs is the subject of many theoretical investigations discussing band structure topology \cite{RN25}, possible pairing states \cite{RN86, RN3}, spin orbit coupling \cite{RN85}, topological properties \cite{RN89}, local non-centrosymmetry \cite{RN83}, and gap function symmetry analysis \cite{RN84}, there are very few experimental reports, all of which are limited to polycrystalline samples \cite{RN80, RN9, RN87, RN89, RN93}. Several previous works mention the advantages of growing single crystals of SrPtAs to further probe the bulk properties \cite{RN25, RN87, RN3, RN88, RN93, RN89}. To this end we set out to synthesize and characterize single crystals of SrPtAs.
\begin{figure}[!ht]
    \begin{center}
    \includegraphics[width=0.5\columnwidth]{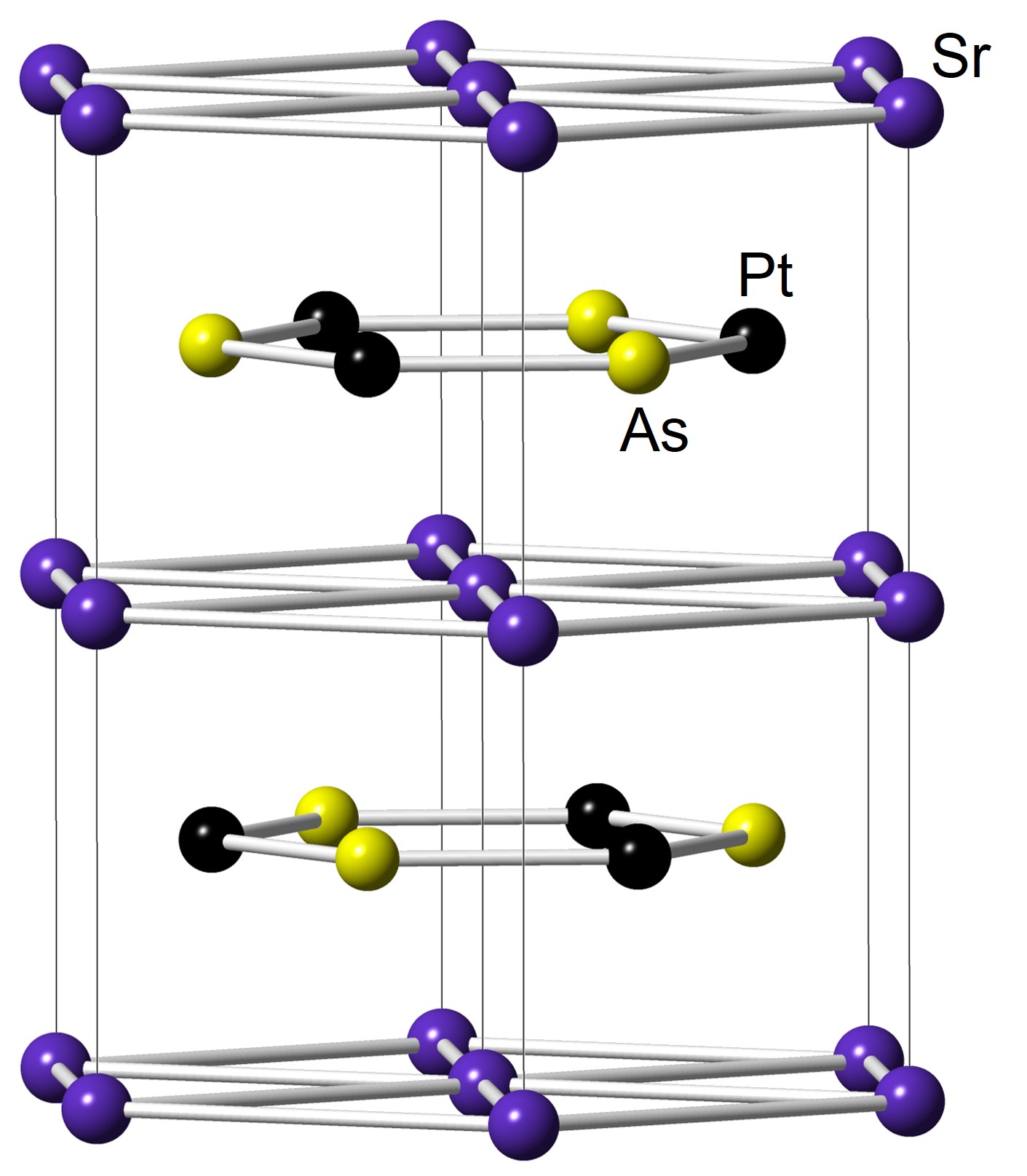}
    \vspace{-0.5cm}
    \end{center}
    \caption{Crystal structure of SrPtAs (\textit{P}6$_{3}$/\textit{mmc}, No. 194), with Sr in purple, Pt in black, and As in yellow.}
    \label{figure:Fig1}
\end{figure}
This work presents the synthesis and physical properties of single crystalline SrPtAs. Although magnetic susceptibility and electrical resistivity measurements reveal superconducting transitions at 2.2 K, specific heat and thermal expansion measurements show a significantly decreased T$_c$ or no T$_c$, respectively. Herein, we discuss the varying superconducting transition temperatures and draw parallels to other AlB$_2$ derived superconductor MgB$_2$.

\begingroup

\begin{table}[!ht]
    \caption{Crystallographic data, data collection, and refinement parameters of SrPtAs.} 
\begin{tabular}{l@{\hspace{1.1 cm}}c@{\hspace{0.8cm}}}	
\hline
Formula &SrPt$_{0.97}$As$_{1.03}$\\
Space Group &\textit{P}6$_{3}$/\textit{mmc}\\
\hline
\textit{a} ({\AA}) &4.2445(4)\\	
\textit{c} ({\AA}) &8.9513(18) \\
\textit{V} ({\AA}$^3$)	&139.66(4)	 \\	
Z	&2\\
T (K)	&298\\	
$\theta$ range (degrees) &4.55-30.40\\	
$\mu$ (absorption coefficient, $mm^{-1}$) &79.37	\\	
measured reflections	&6210\\	
independent reflections	&105 \\	
$\Delta\rho_{max}$ (largest peak, e{\AA}$^{-3}$) &2.00\\
$\Delta\rho_{min}$ (deepest hole, e{\AA}$^{-3}$)	&-2.34\\
extinction coefficient &0.031(7)\\
\textit{R}$_1$(\textit{F}$^2$$>$2$\sigma$(\textit{F}$^2$))	&0.029\\
\textit{wR}$_2$(\textit{F}$^2$)	&0.059\\
\hline

\end{tabular}
\end{table}
\endgroup
\begin{table*}[!ht]
    \caption{Fractional atomic coordinates and anisotropic displacement parameters of SrPtAs.}
\centering
\renewcommand{\arraystretch}{1.25}
\begin{tabular}{c@{\hspace{0.25cm}}c@{\hspace{0.4cm}}l@{\hspace{0.4cm}}l@{\hspace{0.4cm}}l@{\hspace{0.4cm}}l@{\hspace{0.4cm}}l@{\hspace{0.4cm}}l@{\hspace{0.4cm}}l@{\hspace{0.4cm}}l@{\hspace{0.4cm}}l@{\hspace{0.4cm}}}
\hline
Site Label & Wyckoff & \textit{x} &\textit{y} &\textit{z} &\textit{U}$_{eq}$ &Occupancy &\textit{U}$^{11}$ &\textit{U}$^{22}$ & \textit{U}$^{33}$ &\textit{U}$^{12}$\\

\hline
Sr1	&2\textit{a}	&0	&0	&0	&0.0099(10) &1  &0.0091(10) &0.0091(10) &0.0117(13) &0.0045(5)\\
Pt2/As2	&2\textit{c}	 &$\frac{1}{3}$	&$\frac{2}{3}$	&$\frac{1}{4}$	&0.0084(4) & 0.85(3)/0.15(3) &0.0069(5) &0.0069(5) &0.0114(6) &0.0034(2)\\
As3/Pt3	&2\textit{d}	&$\frac{2}{3}$	&$\frac{1}{3}$	&$\frac{1}{4}$ &0.0100(9)	&0.884(10)/0.116(10) &0.0070(10) &0.0070(10) &0.0161(12) &0.0035(5)\\
\hline
\end{tabular}
\end{table*}

As described in Methods, single crystals of SrPtAs were grown using Pb flux. SrPtAs, of the KZnAs-type derived from the AlB$_2$ (\textit{P}6/\textit{mmm}) structure type, crystallizes in the hexagonal space group \textit{P}6$_{3}$/\textit{mmc} (No. 194) with lattice parameters \textit{a}~=~4.2445(4)~{\AA} and \textit{c}~=~8.9513(18)~{\AA}, as shown in Figure 1. In the AlB$_2$ structure, the B atoms form honeycomb layers separated by Al atoms in the \textit{c} direction. In contrast, in SrPtAs the B site is occupied by both Pt and As sites, which alternate both within the honeycomb lattice and in the \textit{c}~direction such that above each As atom is a Pt atom and above each Pt atom is an As atom. This deviation from the AlB$_2$ structure results in the doubling of the unit cell along the \textit{c}~axis \cite{RN9} and breaks local centrosymmetry although the global structure remains centrosymmetric. 
SrPtAs consists of three crystallographically unique sites, Sr (2\textit{a}), Pt (2\textit{c}), and As (2\textit{d}), where the Pt and As sites are disordered. A more accurate description therefore is Sr(Pt$_{1-x}$As$_x$)(As$_{1-y}$Pt$_y$) where \textit{x} = 0.15(3) and $y$ = 0.116(10). Table~1 shows data collection and refinement parameters, and Table~2 presents fractional atomic coordinates and displacement parameters.

\begin{figure}[!ht]
    \begin{center}
    \includegraphics[width=1\columnwidth]{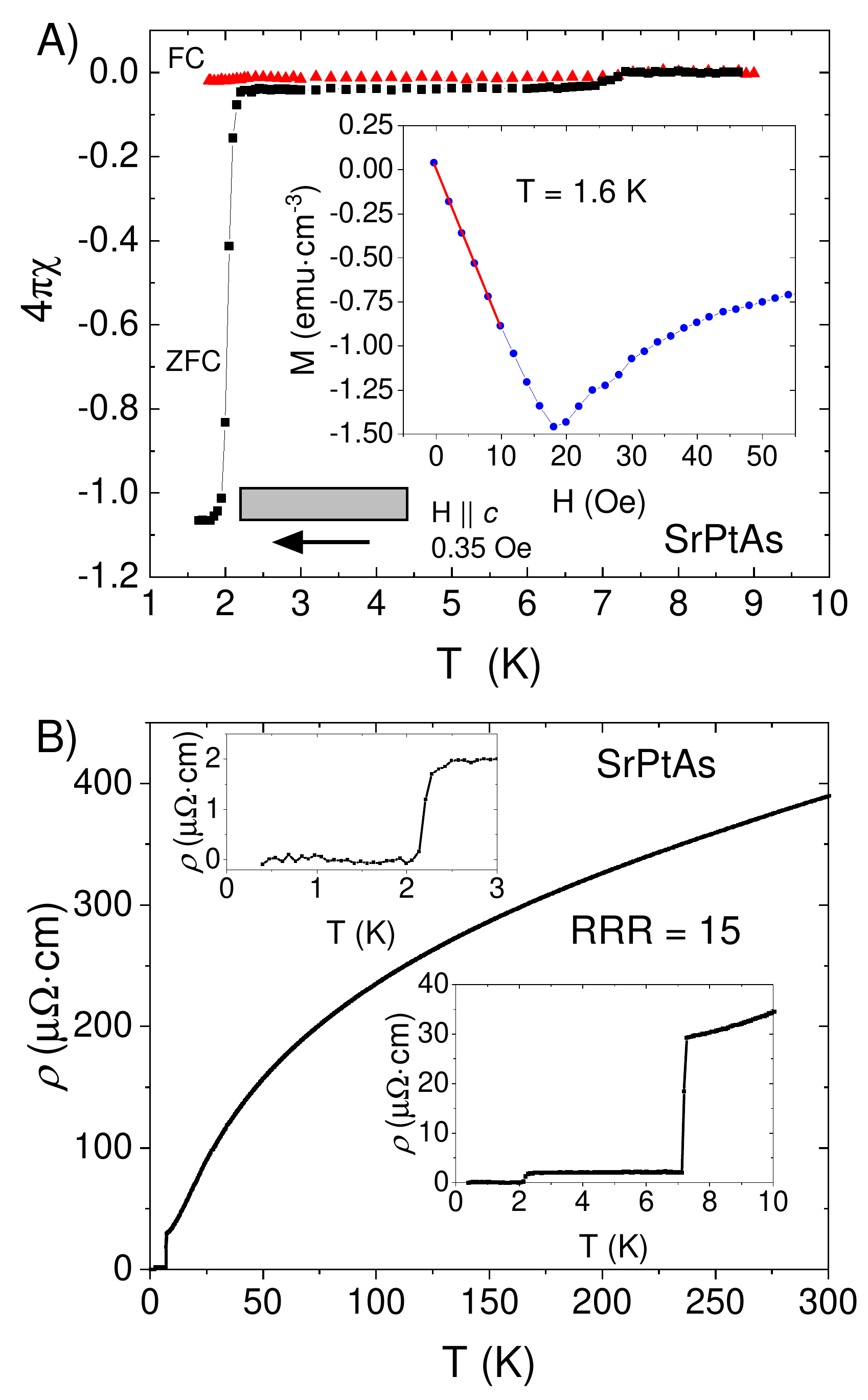}
    \vspace{-0.5cm}
    \end{center}
    \caption{A) Temperature dependent magnetic susceptibility ($\chi$(T)) of SrPtAs with field (H = 0.35 Oe) applied along the \textit{c}~axis. Field cooled (FC) data are shown with red triangles and zero field cooled (ZFC) data are shown with black squares. Inset shows field dependent magnetization (M(H) at T~=~1.6~K) in blue circles with the initial slope denoted with a red line.
    B) Electrical resistivity ($\rho$(T)) of SrPtAs at H~=~0. The bottom right inset shows the low temperature region (below 10~K), highlighting the superconducting transition of Pb at $\sim$7~K. The top left inset shows resistivity below 3~K where the superconducting transition of SrPtAs occurs at 2.2~K.}
\end{figure}

\begin{figure*}[!ht]
    \begin{center}
    \includegraphics[width=\textwidth]{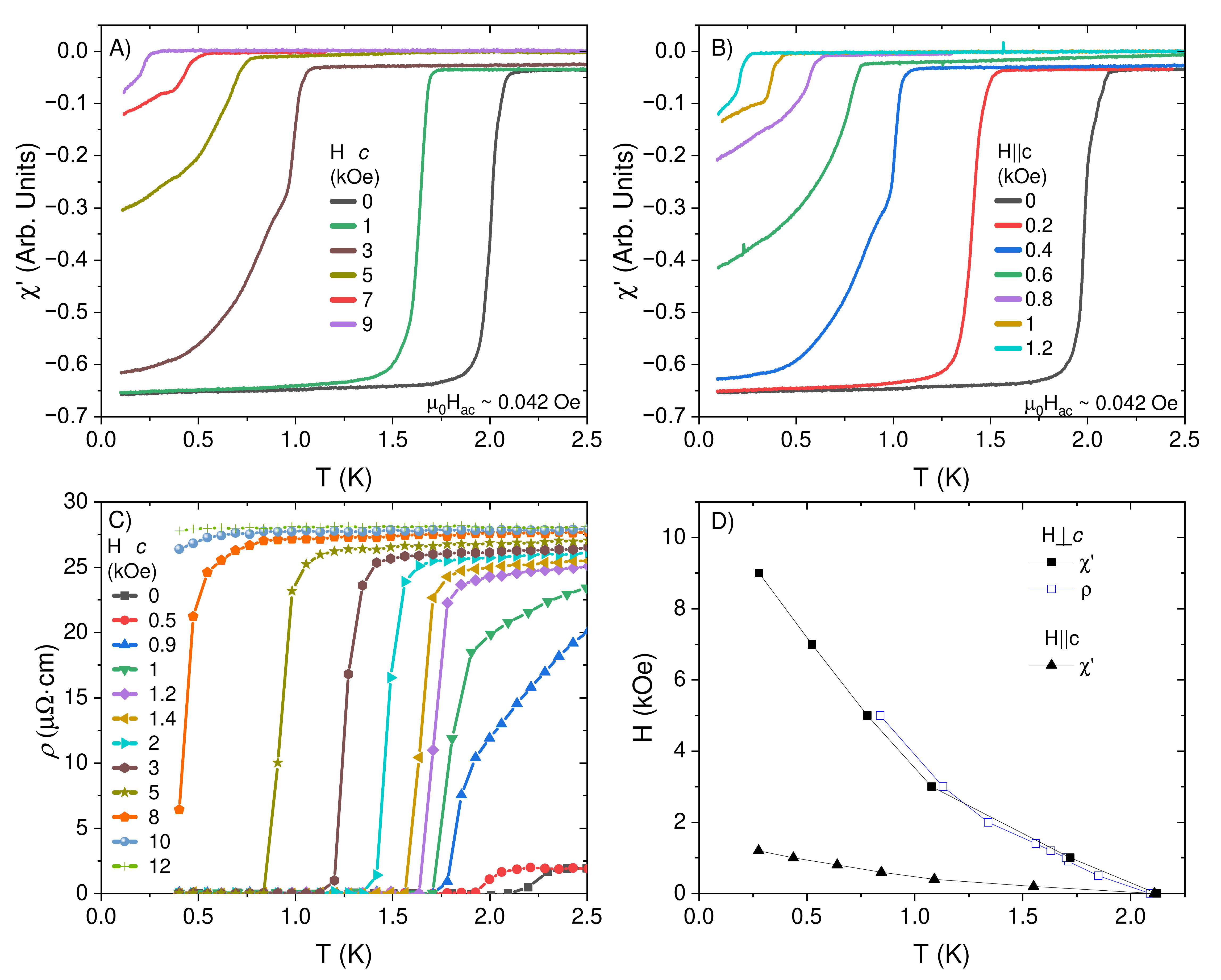}
    \vspace{-0.8cm}
    \end{center}
    \caption{A) $\chi_{ac}$(T) with various fields applied perpendicular to the \textit{c}-axis where the driving ac field $\sim$0.042~Oe. 
    B) $\chi_{ac}$(T) with various fields applied along the \textit{c}-axis where the driving ac field $\sim$0.042~Oe. 
    C) Temperature dependent resistivity ($\rho$(T)) of SrPtAs with varying fields applied perpendicular to the \textit{c}-axis.
    D) Temperature dependence of the upper critical field, $H_{c2}$. Data determined from ac susceptibility (resistivity) measurements are shown in closed black (open blue) symbols. Data collected with field applied parallel (perpendicular) to the \textit{c}~axis are shown with triangles (squares).}
\end{figure*}

The temperature dependent magnetic susceptibility ($\chi$) and electrical resistivity ($\rho$) measurements show a superconducting transition temperature of about 2.2~K, as shown in Figures 2A and 2B, respectively, consistent with previously reported works on polycrystalline samples of SrPtAs \cite{RN80}. Transitions visible at 7 K are due to excess elemental Pb that was used as a flux for the crystal growth.
The susceptibility of SrPtAs with magnetic field (H) (0.35~Oe) applied along the \textit{c}~axis under zero-field cooling (ZFC) and field-cooling (FC) conditions is shown in Figure 2A. A large diamagnetic repulsion response indicates a T$_c$ of $\sim$2.0 K as determined by the midpoint of the drop. 
The ZFC signal, including a demagnetization factor \cite{kittel}, at 1.8~K corresponds to $\sim$~110$\%$ perfect diamagnetism, where $\sim$ 5$\%$ comes from Pb inclusions. A value greater than 100$\%$ could be due to error in the demagnetization factor estimation and/or the presence of chemical impurities. The FC curve has a slight drop that begins at $\sim$~2.2 K. The lack of a large diamagnetic response could be due to nonsuperconducting impurities, vortex pinning, or a lack of bulk superconductivity. The inset to Figure 2A shows the field dependent magnetization (M) wherein the solid red line indicates the initial slope, \mbox{4$\pi\frac{dM}{dH}$~=~-~1.142,} with a corresponding superconducting volume of $\sim$110$\%$. The lower critical field, $H_{c1}$ $\sim$15 Oe at 1.6 K, is estimated from the deviation of the data compared to the initial slope.

The temperature dependent electrical resistivity, $\rho$(T), for a SrPtAs crystal is shown in Figure 2B. The Pb superconducting transition at $\sim$7~K is shown in the bottom right inset. The top left inset shows a superconducting transition at 2.2~K, attributed previously to the bulk superconductivity of polycrystalline SrPtAs. Here, T$_c$ is defined as the midpoint of the resistivity drop. The residual resistivity ratio (RRR), $\frac{\rho_{300K} - \rho_0}{\rho_0}$, is $\sim$15, where $\rho_0$ is determined to be to be $\sim$25~ $\mu \Omega \cdot$cm by extrapolating $\rho(T)$ from above the Pb transition at 7 K to zero temperature. 

The temperature dependent real part of the ac susceptibility ($\chi_{ac}$), collected with various fields perpendicular and parallel to the \textit{c}~axis, are shown in Figure~3A and Figure 3B, respectively. 
The temperature dependence of the electrical resistivity (I$||$\textit{c}) at different fields perpendicular to the \textit{c}~axis is shown in Figure~3C. 
The upper critical field, H$_{c2}$, for both directions, is shown in Figure~3D. T$_c$s for $\chi_{ac}$ were determined by the onset of each transition shown in Figure 3A, and 3B. T$_c$s for resistivity data were determined by the point where $\rho$ = 0 in Figure 3C. For fields parallel (perpendicular) to the \textit{c}~axis, the estimated upper critical field is 1.5 kOe (11 kOe). Similar extrapolations from powder samples of \mbox{SrPtAs} yield an upper critical field ($H_{c2(0)}$) of 2.2~kOe~\cite{RN80}. Utilizing both the isotropic single band Eliashberg model \cite{Eliashberg_C} and Werthamer–Helfand–Hohenberg (WHH) theory \cite{WHH}, $H_{c2(0)}$ was previously calculated (based on polycrystalline SrPtAs) \cite{RN25}. Both values, 1.4 kOe and 1.58 kOe, respectively, agree with our linearly extrapolated lower $H_{c2(0)}$, but are significantly smaller than the $H_{c2(0)}$ perpendicular to the \textit{c}~axis. To calculate the WHH H$_{c2}$ from the anisotropic data of this work, the WHH equation \mbox{$H_{c2}(0)$ = -0.69$T_c(dH_{c2}/dT)_{T_c}$} was utilized, where the resistive $T_c$ of 2.2 K was used. 

For the ac susceptibility data, the resulting H$_{c2}(0)$ for field parallel to $c$ is 0.60 kOe and 4.4 kOe for field perpendicular to $c$. For the resistivity data, H$_{c2}$ for field perpendicular to $c$ is 3.6 kOe. Reasonably, the isotropic value determined from polycrystalline data is between the values for fields parallel and perpendicular to $c$. 
The weak BCS coupling Pauli limit H$_{c2}(0)$ = (1.84(T)*T$_c$(K)) is determined to be 4.0 T (40 kOe) which far exceeds any estimate or extrapolation of H$_{c2}$ for the single crystalline data, possibly suggesting a singlet state.

First-principles calculations \cite{RN85} obtain a quasi-two-dimensional Fermi surface with small corrugations along the $c$ direction. Therefore, the upper critical field along this direction is significantly smaller than $H_{c2(0)}$ perpendicular to the \textit{c}~axis. This, coupled with the upward curvature of the upper critical field curves, could indicate that a multiband model may be applicable.

Multiple aspects of SrPtAs, such as resistivity, ac susceptibility, and the resulting upper critical field curves, are reminiscent of MgB$_2$. For example, (i) the highly anisotropic nature of H$_{c2}$ is reminiscent of MgB$_2$ where H$_{c2}^{c} \sim$~2.5~T and H$_{c2}^{ab} \sim$~16~T resulting in a low temperature anisotropy value of H$_{c2}^{ab}$/H$_{c2}^{c}$ = 6-7 \cite{RN66}, (ii) the upward curvature of H$_{c2}$,  is an indicator of multiband superconductivity \cite{RN65}, and (iii) the ac susceptibility curves exhibit a kink below the onset of T$_c$ \cite{RN68}, (for example, in Figure 4A, at 5 kOe there are two kinks, one at $\sim$1 K and one at $\sim$0.8 K) although less pronounced in the $ab$~plane, which could indicate multiband superconductivity \cite{RN66}, vortex physics \cite{RN66}, surface superconductivity \cite{RN67}, or imperfect crystallinity \cite{RN68}. However, the kinks have the same field dependence and anisotropy as T$_c$. This could indicate a change in the gap structure or vortex excitations.

\begin{figure}[!hb]
    \begin{center}
    \includegraphics[width=1\columnwidth]{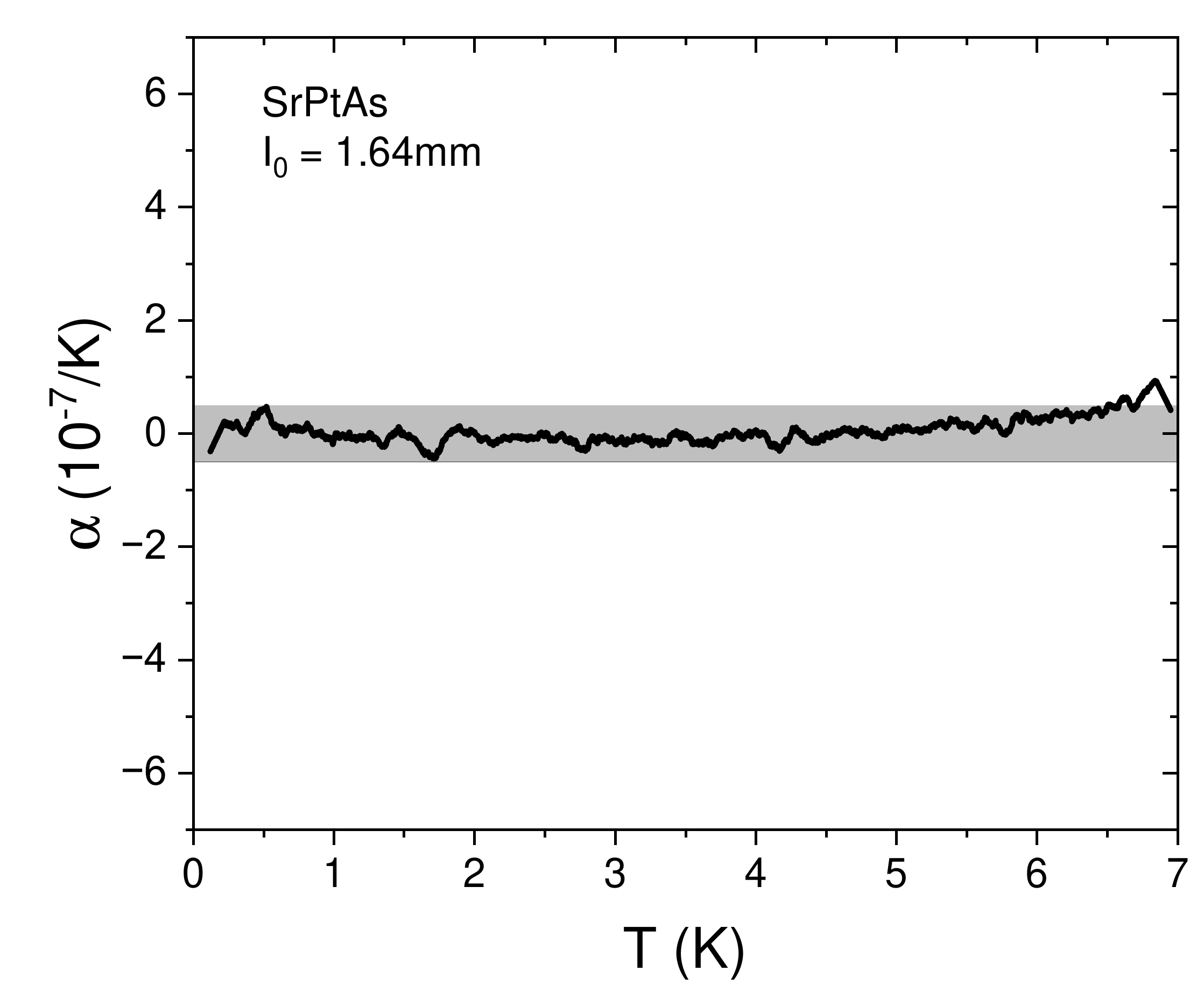}
    \vspace{-0.5cm}
    \end{center}
    \caption{Low temperature thermal expansion of SrPtAs along the \textit{c}-axis from 0.1~K to 7~K (black line). The shaded region indicates the detection threshold of the experimental setup.}
\end{figure}

Thermal expansion of the same SrPtAs crystal along the \textit{c}~axis, shown in black in Figure 4, does not reveal a bulk superconducting transition. Multiple runs reveal statistical noise with no transition. The gray shaded region in Figure 4 indicates the resolution of the experimental setup. The lack of a feature indicating a bulk superconducting transition could be because the magnitude of the transition is below the detection threshold of this experimental setup. Elemental Nb for example has a thermal expansion on the order of 10$^{-8}$K$^{-1}$ which would not be distinguishable in this setup \cite{Nb}. Additionally, SrPtAs may not have much pressure dependence of T$_c$ along the $c$ axis. 

Heat capacity measurements were performed on the same crystal. Figure~5A shows specific heat divided by temperature (C$_p$/T) as a function of temperature. Figure~5B shows C$_p$/T as a function of temperature squared where the red line denotes the fit to the data by C/T~=~$\gamma$~+~$\beta$T$^2$. Here $\gamma$, the electronic specific heat coefficient, is 4.27 mJ/K$^{2}$mol and $\beta$, a constant corresponding to the Debye phonon contribution, is 0.65 mJ/K$^4$mol. The Debye temperature $\Theta_D$ is calculated from:
${\Theta_D}$ = $\sqrt[3]{\frac{12\pi^4nR}{5\beta}}$ where R is the ideal gas constant (8.314 J/molK) and $n$ = 3, the number of atoms per formula unit \cite{Ashcroft76}, yielding $\Theta_D$ = 208 K. 
A small heat capacity jump ($\Delta C/T_c \sim$2 mJ/K$^2$mol) is visible at 0.76 K. Using the value of $\gamma$ from the fit (= 4.27 mJ/K$^{2}$mol), the quantity $\Delta$C/$\gamma T_c$  = 0.47 is obtained. Assuming a weak-coupling value of $\Delta$C/$\gamma T_c$ = 1.43 \cite{tinkham}, a superconducting volume fraction is estimated to be 33\%. 
These values may be compared to $\gamma$ = 7.31 mJ/K$^2$mol, ${\Theta_D}$ = 241 K, and $\Delta$C/$\gamma T_c$ = 1.07 for unpublished SrPtAs data mentioned in \cite{RN61} and $\Delta$C/$\gamma T_c$ for MgB$_2$ of 1.09 \cite{RN90}. 

The discrepancy between the higher T$_c$ values ($\sim$2.0-2.4 K) determined by non thermodynamic measurements compared to the lower T$_c$ ($\sim$0.76 K) bulk measurements continues to be a puzzle. Possible explanations include strain induced filamentary or surface superconductivity and/or the presence of an impurity phase. For example, in CeIrIn$_5$ resistance goes to zero at $\sim$1~K whereas specific heat data indicate bulk superconductivity at 0.4~K \cite{RN53}. This discrepancy is considered to be due to strain introduced by crystallographic defects \cite{RN64, RN366}.  Other examples of filamentary or surface superconductivity include CePt$_3$Si \cite{RN75} and WO$_{2.90}$ \cite{RN33}. Additional measurements, such as comparisons of multiple growth batches from various different research groups, are required to elucidate the underlying reason for the discrepancy in T$_c$. Because the recipe for the growth of single crystals has been determined the opportunity for additional measurements on single crystals of SrPtAs is now available.

\begin{figure}[!ht]
    \begin{center}
    \includegraphics[width=1\columnwidth]{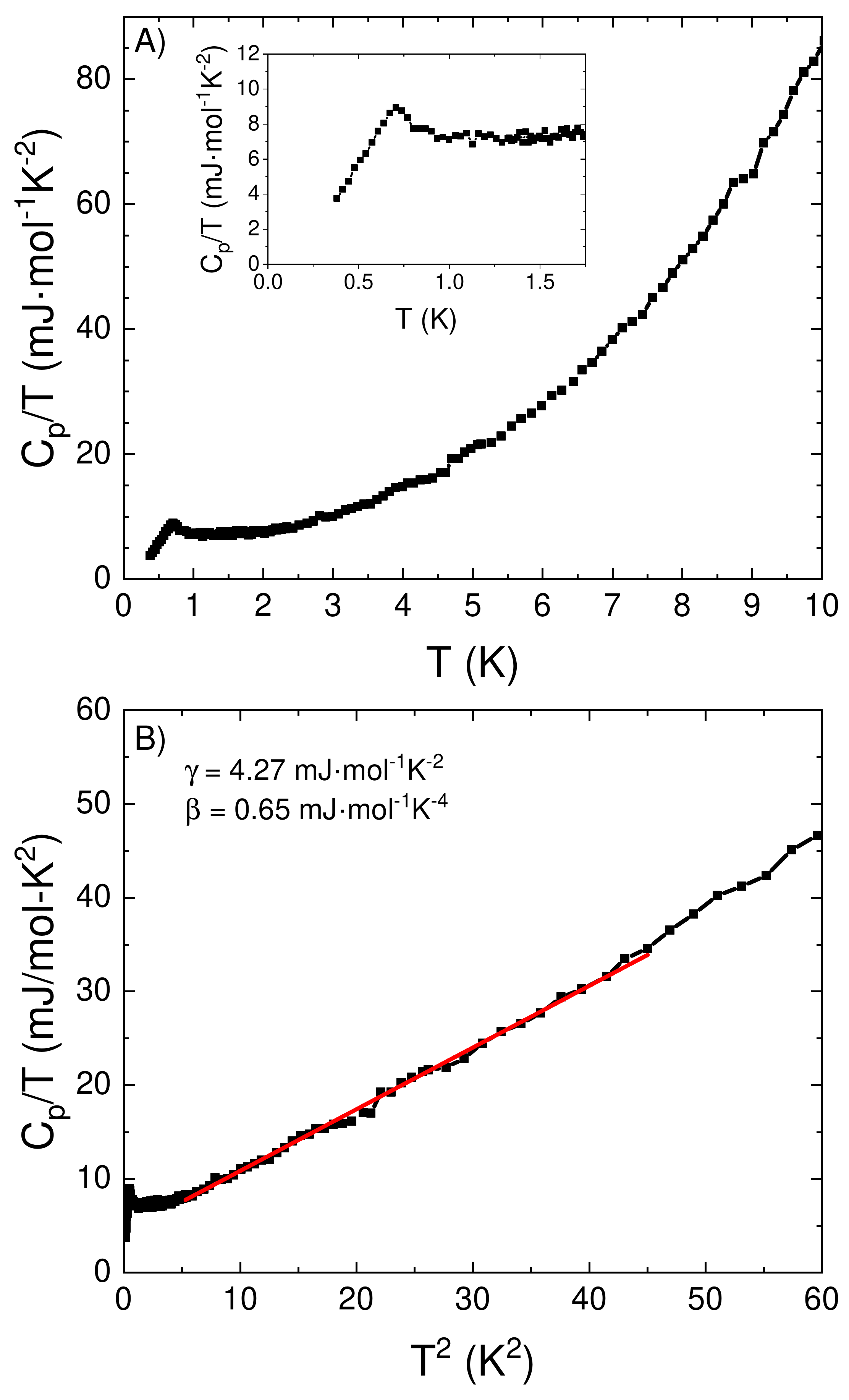}
    \vspace{-0.5cm}
    \end{center}
    \caption{A) Heat capacity divided by temperature (C/T) shown as a function of temperature for SrPtAs. Inset shows zoomed low temperature region. B) C$_p$/T as a function of T$^2$. The red line denotes a fit to the data of C/T~=~$\gamma$~+~$\beta$T$^2$.}
\end{figure}

In summary, we have presented the synthesis and characterization of single crystalline SrPtAs. Magnetic susceptibility and electrical resistivity measurements show superconducting transitions at 2.0-2.4~K, consistent with previous works \cite{RN80, RN87, RN89, RN93}. However, bulk specific heat show decreased T$_c$ of 0.7~K. Multiple similarities between SrPtAs and MgB$_2$ have been discussed and the possibility of SrPtAs as a multiband superconductor has been considered. Our results on single crystals provide a new avenue to explore the underlying mechanisms for the superconductivity of SrPtAs.

\subsection{Methods}
Crystals of SrPtAs were grown in a Pb flux with a reaction ratio of 1:1:0.7:20~of~Sr:Pt:As:Pb. 
The elements were weighed and placed into an alumina crucible set \cite{RN2}.
The crucibles were placed in a fused-silica tube and flame sealed under vacuum using a hydrogen torch.
The reaction vessel was then heated in a programmable furnace to 1150~$^{\circ}$C in 72~hours, 
held for 72~hours, then cooled to 600~$^{\circ}$C at a rate of 3~$^{\circ}$C/h.
The vessel was removed from the furnace at 600~$^{\circ}$C and inverted into a centrifuge to decant the Pb flux.
Reaction ratios of 1:1:1:20 of Sr:Pt:As:Pb following the same heating profile 
resulted in crystals of non superconducting SrPt$_{0.7}$As$_{0.9}$ crystallizing in the ${P}$$\overline{6}$${m}$2 space group.
A single crystal $<$ 0.1 mm on each edge was cut from the larger needle used for property measurements and mounted 
to a \mbox{CryoLoop} using vacuum grease. 
Data were collected with a Bruker~D8 Venture single crystal X-ray diffractometer with an Incoatec I$\mu$S microfocus source (Mo~radiation $\lambda$~=~0.71073~\AA) 
and a PHOTON~II~CPAD area detector. 
The raw data frames were processed with the Bruker SAINT software, and a multi-scan absorption correction was applied with Bruker SADABS. \cite{Krause:aj5242} 
Starting crystallographic models were obtained in SHELXT \cite{Sheldrick:sc5086} using the intrinsic phasing method, and least-squares refinements were performed with SHELXL2018. \cite{Sheldrick:fa3356}
Single crystals of SrPtAs were kept in an argon glovebox between measurements. 
Magnetization measurements were obtained through a commercial Quantum Design MPMS SQUID-based magnetometer. 
Specific heat measurements were made using a commercial Quantum Design PPMS calorimeter that utilizes a quasi-adiabatic thermal relaxation technique. 
The electrical resistivity $\rho$ was characterized using a standard four-probe configuration with an ac resistance bridge. 
Thermal expansion measurements were performed using a capacitance dilatometer described in Ref. \cite{RN1} in an adiabatic demagnetization cryostat.
ac susceptibility was measured using a set of commercial drive and pickup coils. The ac excitation field was estimated to be 0.042 Oe based on the geometry of the coils. 

\begin{acknowledgments}
The work at Los Alamos National Laboratory was primarily supported by the U.S. Department of Energy, Office of Science, National Quantum Information Science Research Centers, Quantum Science Center.
Scanning electron microscope and energy dispersive X-ray measurements were supported by the Center for Integrated Nanotechnologies, an Office of Science User Facility operated for the U.S. Department of Energy Office of Science.
EDB acknowledges support from the U.S. Department of Energy, Office of Basic Energy Sciences, Division of Materials Sciences and Engineering, under the "Quantum Fluctuations in Narrow Band Systems."
AW acknowledges support from the Laboratory Directed Research and Development program at LANL.
\end{acknowledgments}

%

\end{document}